\documentclass[fleqn]{myarticle}
\newfont{\mainfont}{cmss12 scaled 1100}      %font for maintext
\newfont{\titlefont}{cmss12 scaled 1600}   %Title font
\newfont{\authurfont}{cmcsc10 scaled 1000}   %Author name font
\newfont{\authuraddfont}{eurm10 scaled 1000} %Author address font
\newfont{\Labelfont}{cmss12 scaled 1200}   %Section title font. 
\newfont{\labelfont}{cmss12 scaled 1100}   %Subsection title font
\newfont{\figfont}{cmss10 scaled 1000}       %Figure font
\newfont{\footfont}{cmss10 scaled 1000}      %footnote font
\newfont{\tabfont}{cmss10 scaled 1000}       %Table font
\newfont{\reffont}{cmss10 scaled 1000}       %font for reference.
\newfont{\refnamefont}{cmss12 scaled 1100} %font for the label Reference.
\newfont{\abstractfont}{cmss12 scaled 1000}  %font for abstract
\newfont{\abstractnamefont}{eurm10 scaled 1200} %font for the name of Abstract
\usepackage{amssymb}
\usepackage{epsf}
\usepackage{epsfig}
    \newcommand{\ba}{\begin{eqnarray}}
    \newcommand{\ea}{\end{eqnarray}}
    \newcommand{\be}{\begin{equation}}
    \newcommand{\ee}{\end{equation}}

    \newcommand{\phd}{\phi^{\dagger}}
\newcommand{\AmS}{{\protect\the\textfont2
  A\kern-.1667em\lower.5ex\hbox{M}\kern-.125emS}}

\title{ \vspace{-2.0cm} \hfill \parbox{50mm}
{ \large CERN-TH/97-207 }\\
\vspace{2.0cm}
Liapunov Exponents and the Reversibility of 
Molecular Dynamics Algorithms}

\author{
            {\bf Chuan Liu}\\ 
            Department of Physics\\
            Peking University\\
            Beijing, 100871, P.R. China
\vspace{0.2cm} \\
            {\bf Andreas Jaster}\\ 
            Institut f. Theoretische Physik \\
            der Technischen Universit\"at Braunschweig \\
            Mendelssohnstr. 3  \\
            D-38106 Braunschweig, Germany 
\vspace{0.2cm} \\
       {\bf Karl Jansen} \\
        CERN--Theory Division \\
       CH-1211 Gen\`eve 23, Switzerland  
         }
       
\begin{document}

\maketitle

\begin{abstract}
%\settowidth{\baselineskip}{\abstractfont QC}
\setlength{\baselineskip}{0.65cm}
\abstractfont{
We study the phenomenon of lack of reversibility 
in molecular dynamics algorithms
for the case of Wilson's lattice QCD.
We demonstrate that the classical equations of motion that are
employed in these algorithms are chaotic in nature.
The leading Liapunov exponent is determined in a range of coupling
parameters. We give 
a quantitative estimate of the
consequences of the  breakdown of reversibility due
to round-off errors.} 
\end{abstract} 

\newpage

\mainfont       %%% set font for the document

\section{Introduction}

Ever since Wilson's formulation of QCD for a Euclidean lattice,  
numerical simulations have played an important role in 
non-perturbative investigations of physics concerning
the strong interaction. When performing numerical simulations of lattice
QCD, especially with dynamical fermions, 
one usually relies on molecular dynamics kind of algorithms 
such as the Hybrid Monte Carlo (HMC) \cite{tony} --
the most popular one -- or the Kramers equation algorithm 
 \cite{horowitz,kra_us}.  

A major ingredient of these algorithms is the
--numerical-- integration of a set of non-linear differential
equations, the classical equations of motion. 
The non-linearity of these equations may give rise to the suspicion that the
equations of motion may behave like those of a chaotic classical dynamical system. 
Indeed, it was first shown in \cite{kra_us} and later confirmed 
in \cite{wuppertal,brower,ivan} that the equations of motion 
used in the HMC algorithm, when applied to QCD, are chaotic in nature. 
This amusing but, at first sight, academic observation has an important
consequence for any practical application of such kind of algorithms. 

The exactness of 
the molecular dynamics kind of algorithms is guaranteed by the --sufficient-- 
detailed balance condition. 
For this condition to hold, 
the equations of motion for the system, which govern the classical motion, 
are to be reversible.
However, as noticed some time ago \cite{bob}, this reversibility
condition is violated due to round-off errors occurring in
the numerical integration of the equations of motion. 
The fact that the equations of motion are chaotic in nature now
means nothing else but that 
the round-off
errors  
get magnified exponentially along the integration of the equations of motion with a
positive Liapunov exponent $\nu$. 
This by now well-established phenomenon \cite{kra_us,wuppertal,brower,ivan,revus} 
therefore leads to 
rounding-error effects much larger than might naively be expected.
Because of these violations of the reversibility condition, 
the algorithms are no longer exact from a principle point of
view. Of course, the question remains at what quantitative level
the reversibility violation manifests itself in physical
observables. 

In this paper, we provide a detailed study of the Liapunov
exponents for the case of QCD 
and determine them in a range of values for the couplings of the
theory. A short account of our results has appeared in \cite{revus}. 
The dependence of the leading Liapunov exponent 
on the coupling constants 
allows us to estimate 
where, in the coupling parameter
space, we may expect a stronger or weaker exponential amplification
of rounding-error effects. We then try to estimate how much the 
reversibility violation affects physical observables such as the
plaquette, the Polyakov line, plaquette correlation functions 
and the scalar density. 
Finally we emphasize the special role of the acceptance
Hamiltonian with respect to the rounding-error effects.
 
Although we will discuss these effects here for the
case of lattice QCD, we think that similar problems may also arise
in different fields where molecular dynamics algorithms are used, 
e.g.\ in fluid dynamics. 
There again it might be that the equations of motion used resemble
those of a chaotic, classical, dynamical system with a positive 
Liapunov exponent. 
In some cases, as in part of the QCD simulations presently done, 
computers with 32-bit arithmetic are used. For these case, especially, 
it is  
desirable to know whether rounding-errors have a noticeable
and significant 
influence on the values of physical observables. 
In addition, to speed up the numerical simulations, 
even on 64-bit arithmetic machines the time-consuming 
part of the computational work is sometimes performed
in 32-bit arithmetic, for which case one would again be interested
in the rounding-error effects. 

\section{Molecular dynamics equations for lattice QCD}
\label{sec:hmc}

Let us briefly sketch the HMC algorithm
that is used for simulations of lattice QCD. 
Consider first for simplicity a quantum mechanical system 
quantized by Feynman's path integral prescription and specified by its action $S(q)$,
where $q$ denotes a path of the quantum mechanical particle.
In order to evaluate Feynman's path integral numerically,  
one sets the system on a Euclidean discrete time lattice. 
The task is then to generate 
{\em classical} paths $q$ that are distributed according
to the Boltzmann distribution $e^{-S}$. 

One way to proceed is to use a method that resembles those used
in molecular dynamics simulations. 
To this end one introduces an additional, fictitious so-called 
Monte Carlo time $\tau$ and corresponding momenta $p$. The HMC
algorithm then works as follows.

One starts from some given path $q_i$ on the discrete time lattice 
and generates  
initial momenta 
$p_i$ 
from a Gaussian distribution of unit variance and zero mean. 
From this initial set $(q_i,p_i)$ one computes an initial classical Hamiltonian 
$H(p_i,q_i)$, where $H=\frac{1}{2} p^2 + S(q)$. The 
fictitious time evolution of the 
momenta and the paths is now given by 
the following set of coupled
first-order differential equations: 
\begin{equation} \label{eqsofmo}
\dot p = -\frac{\delta H}{\delta q} \;,\;\;
\dot q =  p\; .
\end{equation}
The time derivatives in eqs.~(\ref{eqsofmo}) are to be understood with respect to 
the fictitious Monte Carlo time.
The numerical integration of the equations
given in eqs.~(\ref{eqsofmo}) is performed  
by using a discretized form of eqs.~(\ref{eqsofmo}). In practice one uses
a leap-frog integration scheme, using $N_{md}$ integration steps of
size $\delta\tau$ in order to integrate from fictitious time $0$ to some value $\tau$. 
Such an integration procedure is called a trajectory of length $\tau$.   

In order to generate the desired 
Boltzmann 
distribution and to account for the discretization errors of the integration procedure, 
the new momenta $p_f$ and coordinates $q_f$ 
obtained 
after the integration of the equations of motion 
are only accepted with 
a certain probability $P$,
$P=\textrm{min}(1,\exp\left\{-\Delta H\right\})$ 
with $\Delta H = H(p_f,q_f) - H(p_i,q_i)$. 
The Hamiltonians that enter this Metropolis step are called the 
acceptance Hamiltonians. 
The desired distribution is now obtained by repeating this
procedure a sufficiently large number of times.
The above prescription ensures the detailed balance condition, and the resulting 
algorithm is hence exact in the sense that it indeed converges to the
anticipated Boltzmann distribution. 

The heart of the algorithm sketched above is the integration of the 
classical equations of motion eqs.~(\ref{eqsofmo}). This part is reminiscent
of techniques used in molecular dynamics algorithms. Quite often the equations 
of motion derived from the problem under consideration are non-linear and may as such 
correspond to those characterizing a chaotical dynamical system. 

Let us now come to our example of 
lattice QCD, which is a relativistic field theory defined 
in 4-dimensional Euclidean space-time. 
To be specific, we will focus our discussion on
the equations that arise in the simulations of QCD in
Wilson's formulation 
with two flavours of quarks with degenerate masses.

Lattice QCD
 is established on a Euclidean space-time lattice of  
size $N^3\times T$. With lattice spacing set to unity,  a
point $x$ on the lattice has integer coordinates $x=(t,x_1,x_2,x_3)$, which 
are in the range $0\le t < T;0\le x_i < N$. 
A gauge field $U_{x,\mu}\in SU(3)$ is assigned to the link
pointing from point $x$ to point $(x+\mu)$, where 
$\mu=0,1,2,3$ designates the four forward directions in space-time.
In this study, periodic boundary conditions have been taken
for the gauge fields and for the quark fields in all four directions.
The full partition function for Wilson QCD is given by
\be
{\mathcal Z}=\int {\mathcal D}U
       \exp\,(- S_g) \,\textrm{Det}(Q^2)
 =\int {\mathcal D}U{\mathcal D}\phd{\mathcal D}\phi
       \exp\,\left(- S_g-\phd Q^{-2}\phi \right) \;\;.
\ee 
The term $S_g$ in the exponential is the pure gauge action and is given by
\be
\label{eq:gauge}
S_g = -{\beta \over 6} \sum_{P} \textrm{Tr}(U_P+U^{\dagger}_P)\;\;. 
\ee
The symbol $U_P$ represents the usual plaquette term on the lattice.
The determinant factor $\textrm{Det}(Q^2)$ represents the contribution of
internal fermion loops to the theory. 
%In the so-called valence
%(quenched) approximation, this factor is set to unity identically.
The matrix $Q$ that appears in the determinant is a Hermitian
 sparse matrix defined by:
\be
Q(U)_{x,y}=
c_0\gamma_5\left[\delta_{x,y}-
\kappa \sum_{\mu}(1-\gamma_{\mu})U_{x,\mu}\delta_{x+\mu,y}
+(1+\gamma_{\mu})U^{\dagger}_{x-\mu,\mu}\delta_{x-\mu,y}\right]\;\;,
\ee
with $\kappa$  the so-called hopping parameter and
$c_0=1/(1+8\kappa)$.
%The rewriting of the determinant of $Q^2$ in terms of 
%bosonic integration over $\phd$ and $\phi$ is purely for
%the ease of numerical simulation.
The aim of the simulation is to generate configurations according
to the probability distribution 
$\exp(-S_{eff})\equiv\exp(-S_{g}-\phd Q^{-2}\phi)$ 
%which is a complicated function
%of all the gauge fields $U_{x,\mu}$, using Monte Carlo methods.
 using Monte Carlo methods.

The Hamiltonian for the simulations of lattice QCD is given by
%Monte Carlo time \cite{tony,sugar}. For this purpose, 
\be
\label{eq:ham}
{\mathcal H} = \sum_{x,\mu}{1 \over 2}\textrm{Tr}(H^2_{x,\mu}) 
+ S_{eff}(U_{x,\mu},\phd,\phi)\;\;,
\ee
where $H_{x,\mu}$ is the momentum conjugate to the 
gauge field $U_{x,\mu}$ and 
takes values in $su(3)$, the Lie algebra of $SU(3)$. 

Since the bosonic part $S_b\equiv \phd Q^{-2} \phi$ is quadratic 
in the $\phi$ fields, these are generated at the
beginning of each molecular dynamics trajectory via
$\phi= QR$,
where $R$ is a random spinor field which is Gaussian distributed.
The kinetic term in eq.~(\ref{eq:ham}) 
is also generated from Gaussian noise at the
beginning of the update. Then the gauge fields and their
corresponding momenta are updated according to the 
equations of motion:
\be
\label{eq:update_continuum}
 {\dot U}_{x,\mu} =iH_{x,\mu}U_{x,\mu}\;\;, \;\;\; 
i {\dot H}_{x,\mu}= [U_{x,\mu}F_{x,\mu}]_{T.A.} \;\;,
\ee
where the symbol $[\cdots]_{T.A.}$ stands for taking the traceless
anti-Hermitian part of the matrix \cite{sugar} and
the quantity $U_{x,\mu}F_{x,\mu}$ is the total force 
associated with the link $U_{x,\mu}$.
The dot on a field variable represents the 
derivative 
%\footnote{ \footfont These differential equations should
%only be understood in a formal sense in the continuum.}
with respect to the Monte Carlo time $\tau$.
The quantity $F_{x,\mu}$ 
is nothing but the "coefficient" in the change of the
effective action when an infinitesimal change of the
gauge link $\delta U_{x,\mu}$ is applied, i.e.
\be
\label{eq:linear}
\delta S_{eff} = \sum_{x \mu} 
\textrm{Tr}(F_{x,\mu}\delta U_{x,\mu}+F^{\dagger}_{x,\mu}\delta 
U^{\dagger}_{x,\mu})\;\;.
\ee
One can easily check that the time evolution described by
eqs.~(\ref{eq:update_continuum}) conserve 
the Hamiltonian~(\ref{eq:ham}). 
Equations~(\ref{eq:update_continuum}) define a Hamilton
flow in a phase-space manifold that is a direct product of
$4N^3T$ factors of $SU(3)$ and $su(3)$.

Equations~(6) establish the molecular dynamics part of the HMC algorithm and it is
in this part where the problems with rounding-errors appear.
%As mentioned above, the numerical integration of eq.~(\ref{eq:update_continuum}) 
%is performed using a leap-frog method. One usually integrates from 
%Monte Carlo time $\tau =0$ up to $\tau=1$ with $N_{md}$ integration steps of
%step size $\delta\tau$. 
Note that  
%stochastic variables $\eta(\tau)$, which
%in the case of lattice QCD are the 
one considers here
a purely classical set of autonomous first-order differential equations. 

\section{Liapunov exponents}
\label{sec:exponents}

In this section, we briefly describe the concept of the Liapunov
characteristic exponents \cite{liapunov,oseledec,benettin1}.  
%for the equations of motion in
%molecular dynamics algorithms. 
Liapunov exponents serve as an
important quantitative measure for the degree of stochasticity of
a dynamical system. For an introduction to this topic, see \cite{llbook} and
references therein.

Let us consider a time evolution $q(\tau)$, $p(\tau)$ , described 
by a set of
first-order autonomous differential equations like eqs.~(\ref{eqsofmo}). 
To each point in the phase-space manifold $(q,p)$ one can attribute 
a local 
Liapunov exponent.
These exponents describe the mean exponential
rate of divergence of the distance between the 
trajectories of two nearby points in phase-space.
Given two points in the phase-space manifold, which are
close to each other at $\tau=0$, one can follow the time
evolutions of the trajectories originating from these two points. 
At any given instance of the time $\tau$, one can construct 
the vector pointing from a point on one of the trajectories 
to the corresponding one on the other trajectory.
If the distance between the  two initial points 
in phase-space becomes infinitesimally
small, the above-mentioned vector belongs to the tangent
space of the phase-space manifold. 

The Liapunov exponents can now be determined by studying
the time evolution of these tangent vectors.
In order to be specific, we will directly discuss the
example of the equations of motion used in the HMC algorithm
as applied for lattice QCD. 
We start   
denoting by $dH_{x,\mu}$ and $dX_{x,\mu}=-iU^{-1}_{x,\mu}dU_{x,\mu}$
the tangent vectors of $H_{x,\mu}$ and $U_{x,\mu}$, respectively.
The time evolution of these vectors is described by 
\be
\label{eq:update_d}
 {\dot {dX}}_{x,\mu} =U^{-1}_{x,\mu}dH_{x,\mu}U_{x,\mu}\;\;,\;\;\;\; 
i {\dot {dH}}_{x,\mu}= d[U_{x,\mu}F_{x,\mu}]_{T.A.} \;\;,
\ee
which follow from eqs.~(\ref{eq:update_continuum}).
One can study this set of equations for $(dH_{x,\mu},dX_{x,\mu})$ 
together with the original eqs.~(\ref{eq:update_continuum}).
Note that  eqs.~(\ref{eq:update_d}) are
linear in $dH_{x,\mu}$ and $dX_{x,\mu}$, 
both of which are elements of $su(3)$. 
Let us now assume that, initially at
$\tau=0$, $(dH_{x,\mu},dX_{x,\mu})$ 
were given some value in the tangent space of some point in the
phase-space manifold. We can introduce the norm in the
tangent space as:
\be
D^2(\tau)
= \sum_{x,\mu} \textrm{Tr}(dH^2_{x,\mu}(\tau)+dX^2_{x,\mu}(\tau))\;\;.
\ee
As time evolves, this norm will change and the Liapunov exponent
can be defined as:
\be
\nu= \lim_{D(0) \rightarrow 0}\lim_{\tau \rightarrow \infty} 
{1 \over \tau}\log{ D(\tau) \over D(0) }\;\;.
\ee
In fact, due to its linear nature in $(dH_{x,\mu},dX_{x,\mu})$,
 eqs.~(\ref{eq:update_d}) can be written as 
\be
{d \over d\tau} \left(\begin{array}{c}
                  dH_{x,\mu}\\
                  dX_{x,\mu}\\
             \end{array}\right)
= J[H_{x,\mu},U_{x,\mu}] \cdot 
 \left(\begin{array}{c}
                  dH_{x,\mu}\\
                  dX_{x,\mu}\\
             \end{array}\right) \;\;,
\ee
and the Liapunov exponents are just the time averages 
of eigenvalues of the matrix $J$ along a trajectory.
These exponents can be ordered according to the magnitude of
their real part. The one with the largest real part is called the
leading Liapunov exponent and will be studied in more detail below.
It is known \cite{oseledec} that Liapunov exponents do not depend on the
choice of metric in phase-space.

The numerical calculation of Liapunov 
exponents of a given flow can be
done straightforwardly \cite{benettin1}. We now go to
the discretized versions~\footnote{\footfont  
 In the discretized version of numerical
integration, we replace the differentials by the corresponding differences. 
 } 
of eqs.~(\ref{eq:update_continuum}) and eqs.~(\ref{eq:update_d}) 
and integrate them simultaneously in time 
using a leap-frog integration scheme. 
As a safety measure,  
in order to avoid severe rounding-errors,
the tangent vector should be renormalized 
with respect to its initial norm 
after each 
integration step. 
The procedure goes as follows:
Starting at $\tau=0$, we have some initial value of
$D^2(0)=\sum_{x,\mu}\textrm{Tr}(dH^2_{x,\mu}+dX^2_{x,\mu})$ for
a given initial tangent vector $(dH_{x,\mu},dX_{x,\mu})$.
 We then integrate one step in time
with step size $\delta\tau$ using the leap-frog scheme. 
Now, we evaluate the norm
$D(\delta\tau)$ of the new tangent vector and store this information
for the computation of the Liapunov exponents. 
 Next, we rescale the new tangent vector in such a way that
its norm is still equal to $D(0)$, i.e.\
$(\overline{dH}_{x,\mu}(\delta \tau),\overline{dX}_{x,\mu}(\delta \tau))
=(dH_{x,\mu}(\delta \tau),dX_{x,\mu}(\delta \tau))/D(\delta \tau)$.
We then proceed to the next integration step, starting from this
already normalized tangent vector.
Because of the linear nature of eqs.~(\ref{eq:update_d}), this rescaling is
legitimate since the Liapunov exponents only depend on the
ratio of the norms.  We then proceed to
integrate over $n$ such steps. It can be shown \cite{benettin1} that the
average
\be
\nu_n={1 \over n\delta\tau} \sum^{n}_{k=1} \log{D(k\delta\tau) \over D(0)}
\ee
is approaching the leading Liapunov exponent when $n \rightarrow \infty$.
In addition, its value is independent
of the value of $\delta\tau$, as long as the step size
$\delta\tau$ is not too large. 

%Here we would like to pause for a while and point out some
%of the differences between truly classical Hamiltonian
%flows and the molecular dynamics equations used in the
%Monte Carlo simulations. In pure classical Hamilton mechanics, 
%the phase-space manifold is usually partitioned into 
%regular regions and stochastic regions. Different 
%stochastic regions may or may not be connected with one 
%another. Therefore, the value of the Liapunov exponent could  
%depend on where the initial point is in the phase space manifold.
%For example, starting in the regular region of the phase space,
%one would find vanishing value of exponents while in the
%stochastic region, the leading exponent is usually positive.
%The value of the Liapunov exponent for a particular region
%represents the phase space average of the exponential divergence rate in
%that region. 
In contrast to the truly classical Hamilton flows, for which
the phase space is usually partitioned into regular and 
stochastic regions, 
in the case of the HMC algorithm considered here,
the exponent we find is 
a phase space average of local exponents weighted with the appropriate
Boltzmann factor.

Another remark is that the concept of Liapunov exponents could
be generalized to higher-dimensional objects as well; not surprisingly,
the leading exponent for a $p$-dimensional volume formed by
$p$ linear independent tangent vectors is the sum of the
$p$ leading exponents of the tangent vector. We point this out because this
observation can  serve as a tool for calculating the subleading
Liapunov exponents if one wishes to.

\section{Liapunov exponent for simulations of QCD}

We have used the method described in the previous section to
study the Liapunov exponents in simulations of QCD.
A determination of the leading Liapunov exponent
for QCD simulations with gauge group $SU(2)$ was 
reported in \cite{kra_us}. Similar studies were done
for gauge group $SU(3)$ in \cite{wuppertal,ivan}.
 Here, we would like to extend these studies by computing the 
Liapunov exponents in a range of coupling parameter values. 

When numerically integrating the equations 
of motion derived from the lattice QCD action, we have
chosen a step size of $\delta\tau=0.02$, 
which results in an acceptance
of almost $100\%$. We have integrated the equations of motion
with $200$ steps, which amounts to a trajectory length of $4.0$. 
The calculations are mainly done on a $4^4$ lattice but we also used
$8^4$ and $16^4$ lattices. 
We checked that our results 
 do not depend on the value of the step size for
$\delta\tau \leq 0.02$. The initial values of $(dH_{x,\mu},dX_{x,\mu})$
are generated from a Gaussian distribution.
Then the norm of the tangent vector is recorded for each 
time step, from which we could define the ``instantaneous'' 
Liapunov exponent at the $k$-th step as: 
%\be
%\label{eq:ins_expg}
%\nu(\tau) = {1 \over \delta\tau} \log{D(\tau) \over D(0)}\;\;,
%\;\;\;\; \tau=k\delta\tau\;\;.
%\ee
\be
\label{eq:ins_expg}
\nu(\tau) = {1 \over \delta \tau} \log{D(\tau) \over D(0)}\;\;,
\;\;\;\; \tau=k\delta\tau\;\;.
\ee
Note that, as discussed above, 
in the actual simulation $D(\tau)$ is always normalized 
to the norm of the starting vector in order to avoid potential
rounding-errors. 
In order to monitor the change of $dH_{x,\mu}$ with $\tau$ and
 $dX_{x,\mu}$  separately, we have also recorded the
values of $D_H(\tau)$ and $D_X(\tau)$, which are defined as
\be 
D^2_H(\tau)= \sum_{x \mu} \textrm{Tr}(dH^2_{x,\mu}(\tau)) \;\;, \;\;\;\;
D^2_X(\tau)= \sum_{x \mu} \textrm{Tr}(dX^2_{x,\mu}(\tau))\;\;.
\ee
We also define the ``instantaneous'' --or effective-- exponent 
 for $dH_{x,\mu}$ as: 
%\be
%\label{eq:ins_expg_hx}
%\nu_H(\tau) = {1 \over \delta\tau} 
%\log{D_H(\tau) D(\tau-\delta\tau)
% \over D_H(\tau-\delta\tau) D(0)} \;\;,
%\ee
\be
\label{eq:ins_expg_hx}
\nu_H(\tau) = {1 \over \delta\tau} 
\log{D_H(\tau) 
 \over D_H(0)} \;\;,
% \over D_H(\tau-\delta\tau)} \;\;,
\ee
and similarly $\nu_X(\tau)$ for $dX_{x,\mu}$. Again we remark that 
$D_H(\tau)$ receives a proper normalization in the simulation.

We first describe our results for the Liapunov 
exponents obtained in pure $SU(3)$ gauge theory on $4^4$ lattices. 
We have studied the 
theory for various values of  the gauge coupling 
$\beta$ ranging from $\beta=0.5$ up to $\beta=30$. 
%which covers all the
%values of practical Monte Carlo simulations. 
%Then the quantities $\nu(\tau)$, 
%$\nu_H(\tau)$ and $\nu_X(\tau)$ could be
%%%%%%%%%%%%%%%%%%%%%%% Figure for diagrams
\begin{figure}[t]
\vspace{-0mm}
%\centerline{ \epsfysize=7.5cm
%             \epsfxsize=10.5cm
%             \epsfbox{oscillate.ps}}
%\vspace{-0mm}
\begin{center}
\epsfig{%
file=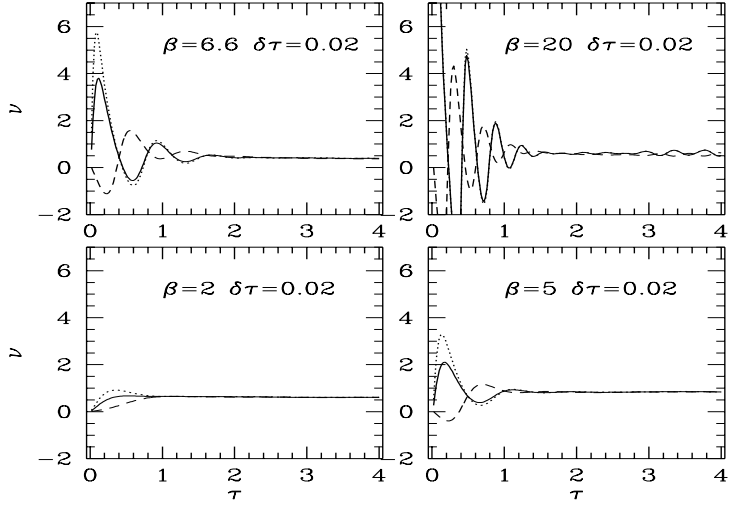,%
height=7.5cm,%
width=10.5cm,%
}
\parbox{12.5cm}{\caption{ \label{fig:ins}
The instantaneous Liapunov exponents, as defined in eq.~(\ref{eq:ins_expg})
and eq.~(\ref{eq:ins_expg_hx}), are plotted as a function of the
Monte Carlo time $\tau$ along a trajectory for various values of $\beta$.
The dotted lines represent $\nu_H(\tau)$ while 
the dashed lines represent $\nu_X(\tau)$.
The $\nu(\tau)$'s are represented by solid lines.
}}
\end{center}
\end{figure}
%%%%%%%%%%%%%%%%%%%%%%% Figure for diagrams
% plotted as a function of $\tau$. 
In fig.~\ref{fig:ins}, we show the 
behaviour of the instantaneous Liapunov exponents
 as a function of the trajectory length for various values of $\beta$. 
From these figures, it is clearly seen that the exponents 
 show an oscillatory behaviour, especially at the
beginning of a trajectory. Then the amplitude of the oscillation
dies out as time evolves. Finally, all three exponents 
$\nu$, $\nu_{H}$ and $\nu_{X}$ approach
the same stable constant value. The leading Liapunov exponent can be
extracted by filtering out the zero frequency part in
$\nu(\tau)$ towards the end of the trajectory. 
In our determination of the leading exponent, we fit the tail of
the function $\nu(\tau)$ to a constant.
% and pick the fitting interval
%by least $\chi^2$ per degree of freedom.
We also note  that as the value of $\beta$ is increased, 
so is the frequency and magnitude of the oscillation.
This is a reflection of the common expectation that for large values of
$\beta$, the system is getting closer to a Gaussian model. 
%%%%%%%%%%%%%%%%%%%%%%% Figure for diagrams
\begin{figure}[t]
\vspace{-0mm}
\centerline{ \epsfysize=7.5cm
             \epsfxsize=10.5cm
             \epsfbox{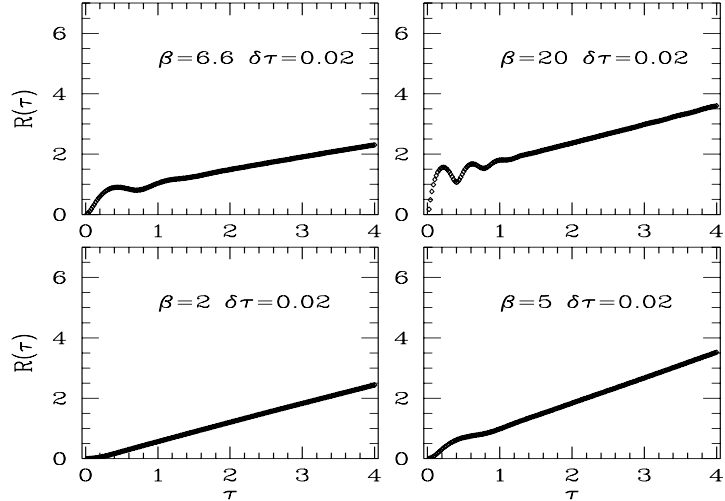}}
\vspace{-0mm}
\begin{center}
\parbox{12.5cm}{\caption{ \label{fig:norm}
The integral of the function $\nu(\tau)$, eq.~(\ref{eq:norm}), which represents 
the change in the norm of the tangent vector as compared
to its original value,
is plotted as a function of the
Monte Carlo time of a trajectory for the same values of $\beta$ as
in fig.~\ref{fig:ins}.
}}
\end{center}
\end{figure}
%%%%%%%%%%%%%%%%%%%%%%% Figure for diagrams
Despite rather strong oscillations for $\nu(\tau)$
 at large values of $\beta$, 
 the integral $R(\tau)$ of the function $\nu(\tau)$, defined as
\be
\label{eq:norm}
R(\tau)=\sum^k_{j=1} \delta\tau \nu(j\delta\tau), \;\;\;\;\; \tau=k\delta\tau ,
\ee
which reflects the change in the norm relative
to the starting point, is basically increasing
linearly with only little oscillations, as shown 
in fig.~\ref{fig:norm}. 

\begin{figure}[t]
\vspace{-0mm}
\centerline{ \epsfysize=7.5cm
             \epsfxsize=10.5cm
             \epsfbox{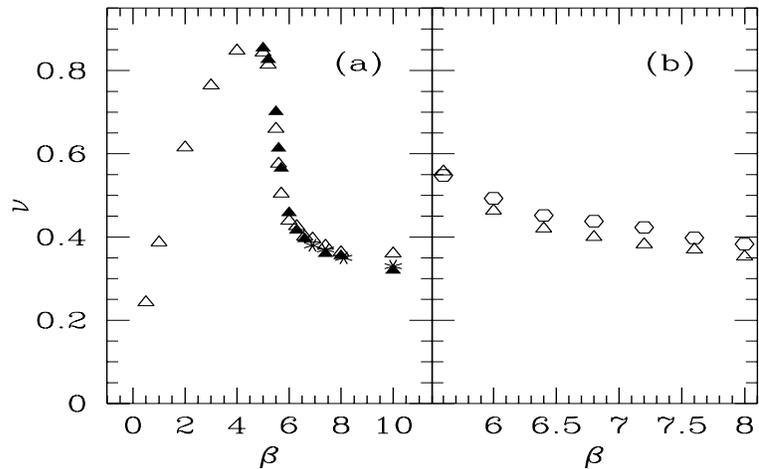}}
\vspace{-10mm}
\begin{center}
\parbox{12.5cm}{\caption{ \label{fig:expg}
The average leading Liapunov exponents for pure $SU(3)$ gauge
theory (a) and full QCD (b) as a function of
 the gauge coupling $\beta$. In (a), 
the open and filled triangles correspond to $4^4$ and $8^4$ lattices,
respectively. The stars denote our results for a $16^4$ lattice. 
In (b), exponents for different values of $\kappa$ are represented by 
different symbols. The triangles correspond to $\kappa=0.13$ and
the hexagons to $\kappa=0.16$. 
}}
\end{center}
\end{figure}

Having extracted the leading Liapunov exponents~\footnote{Similar 
studies have also been done by
the authors of ref. \cite{ivan}. Complete consistent values
of the exponent have been obtained, except for very large $\beta$ values.}
 for several
values of $\beta$, we plot them as a function of $\beta$ in
fig.~\ref{fig:expg} (a). 
%%%%%%%%%%%%%%%%%%%%%%% Figure for diagrams
%%%%%%%%%%%%%%%%%%%%%%% Figure for diagrams
We see from this figure that 
%for very small values of
%$\beta$, the Liapunov exponent is approaching zero. 
%It then has a rather sharp peak around $\beta=3\sim5$.
%At about $\beta=6\sim7$, we observe a sharp drop in the
%exponent.  Then it steadily increases with increasing value of
%$\beta$. 
the Liapunov exponents show a significant $\beta$-dependence. 
%for realistic $\beta$ values in Monte Carlo
From $\beta=5.0$ to  $\beta=10.0$,  
 we have a leading Liapunov exponent that
ranges from $0.85$ to $0.35$.

In fig.~\ref{fig:expg}(a), we plot also results for the Liapunov
exponents for an $8^4$ (filled triangles) and for a $16^4$ lattice
(stars). We can see that the Liapunov exponents for the different size 
lattices fall essentially on one common curve. This behaviour is not in accord 
with 
the anticipated property of the Liapunov exponent to be inversely 
proportional to the correlation length, as proposed in \cite{ivan}.
The data seem to indicate that we are really 
%We can also no theoretical reason,
%why the Liapunov exponent should follow perturbative $\beta$-function of
%SU(3). Remember that we 
studying here classical equations of motion derived from the 
4-dimensional classical Hamiltonian given in eq.~(\ref{eq:ham}).

We have also performed similar investigations of
the Liapunov exponents for
full QCD with dynamical fermions.
 In dynamical QCD simulations, the
parameter $\kappa$ appears and the value
of the Liapunov exponent, in principle, will also depend on 
this parameter. For $\kappa$ values that we have
studied on $4^4$ lattices, 
the leading Liapunov exponent shows, however, only 
a rather weak dependence on $\kappa$.

We have performed a rough scan in the parameter 
ranges $5.6 \leq \beta \leq 8.0$ and 
$0.13 \leq \kappa \leq 0.16$. The Liapunov exponents are measured
in the same way as in the pure gauge case. The instantaneous exponents
also show similar behaviours. In fig.~\ref{fig:expg}(b), 
we show the final 
result of the exponents as a function of $\beta$ and $\kappa$. 
Different $\kappa$ values for the same $\beta$ give 
almost the same value for the Liapunov exponents. 
 For each value
of $\beta$, we have taken four different values of $\kappa$, namely 
$0.13$, $0.14$, $0.15$ and $0.16$. Due to their weak dependence
on $\kappa$, we only show the exponents 
 for two values of $\kappa$, i.e.\ $\kappa=0.13$ (triangles)
and $\kappa=0.16$ (hexagons). 
 The errors of the points are
about the size of the symbols.
In the parameter range that we have studied, the Liapunov exponent
is roughly $0.4-0.5$. 
 
\section{Consequences of irreversibility}

The equations of 
motion in eqs.~(\ref{eq:update_continuum}) are to
be reversible 
to ensure the exactness of
the HMC algorithm used for the
simulations of lattice QCD. 
However, rounding-errors in the numerical integration process of
 this algorithm imply violations of the reversibility condition.  
Thus, the exactness of the algorithm can no longer be guaranteed.         
Although this brings us into an unsatisfactory situation from a
principle point of view, it is necessary to investigate what are 
the consequences of the reversibility violations. 

For a molecular dynamics trajectory there are two causes 
for an increase of rounding-errors when the trajectory length 
$\tau$ is increased for fixed integration step size $\delta\tau$. 
If we assume that the rounding-errors are Gaussian distributed,
we expect a random walk behaviour and hence that the
rounding-errors grow like $\sqrt{\tau}$. Second, as demonstrated
above and also found in \cite{kra_us,wuppertal,brower,ivan}, 
since the equations of motion behave like those of 
a classical chaotic system, the rounding-errors get amplified
exponentially, like $\exp(\nu\tau)$, with a positive Liapunov exponent $\nu$.
 
Both effects lead to a certain amount 
of reversibility violation and in this section we  
try to quantify the effect of this reversibility violation on physical observables 
on the one hand and
on the Hamiltonian that enters the Metropolis decision in the HMC
algorithm on the other hand.

\subsection{Observables} 

To study the effects of the above discussed reversibility violation on observables, 
we have run two
HMC simulation programs in parallel. One of them is with 32-bit
arithmetic but with 64-bit arithmetic scalar products and summations over
the lattice, which mimics the typical situation in simulations on
a 32-bit machine. The other one runs with complete 64-bit arithmetic and 
serves as a reference point for an ``exact'' program. Let us remark that for this
section we have chosen $SU(2)$ as the gauge group. 

We then proceeded as follows: one generates a configuration, say with the 
64-bit arithmetic program version. On this configuration one starts the 32-bit and the 64-bit
program versions and run them for a given number of molecular dynamics steps with
a fixed step size, i.e.\ up to some trajectory length $\tau$. At each step 
of the trajectory,
one measures quantities such as the plaquette $P$, the Polyakov line $L$, plaquette
correlation functions or, in the case of dynamical fermions, the scalar density. 
Let us denote such a quantity with $O$ for the 
64-bit version of the program and with $\tilde{O}$ the 
corresponding one for the
32-bit arithmetic version. 
As an example, let us give the measurement of the 
Polyakov line:
\begin{equation} \label{polyakovline}
L_{\vec{x}}=\frac{1}{2} \textrm{Tr}\,\prod_{t=0}^{T-1} U_{(\vec{x},t),0}\; ,
\end{equation}
i.e.\ the trace of the ordered products of gauge field variables. 
The index $\vec{x}$ denotes the 3-dimensional space coordinate. 
We then measure at each molecular dynamics step 
\begin{equation} \label{localdiff} 
\| L - \tilde{L}\|^2 = \frac{1}{N^3} \sum_{\vec{x}} (L_{\vec{x}}-\tilde{L}_{\vec{x}})^2\; .
\end{equation}
For other observables 
the difference $\| O-\tilde{O}\|$ can be computed in an obviously
generalized analogous way. 
The whole procedure is repeated on a number of configurations in order to 
obtain an error estimate.
From the above discussion we expect that

\begin{equation} \label{difference}
\| O-\tilde{O}\|  = c_0 + c_1\sqrt{\tau} + c_2e^{\nu\tau}\;\; .
\end{equation}

\begin{figure}[htp]
\vspace{-0mm}
\centerline{ \epsfysize=9.5cm
             \epsfxsize=11.5cm
             \epsfbox{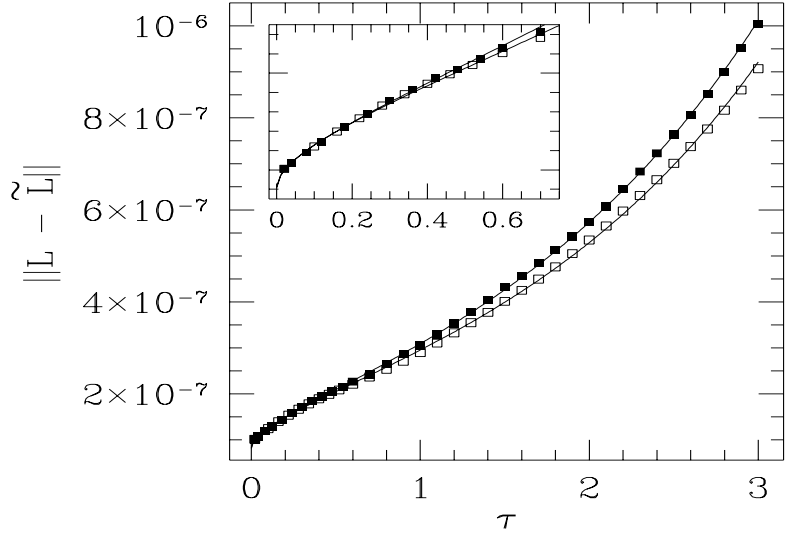}}
\vspace{-0mm}
\begin{center}
\parbox{12.5cm}{\caption{ \label{fig:fit} The difference $\| L-\tilde{L}\|$,
 eq.~(18), 
 for the Polyakov line $L$
 as a function of the trajectory length $\tau$. 
 Open squares are for the pure gauge theory at $\beta=2.12$ 
 and filled squares
 are from full dynamical fermion simulations at $\beta=2.12$, 
 $\kappa =0.15$. The small inset amplifies the region of small
 $\tau$. The solid line is a fit according to eq.~(19).
}}
\end{center}
\end{figure}

The result for the Polyakov line on a $6^38$ lattice is shown in fig.~\ref{fig:fit},
together with a fit to the data according to eq.~(\ref{difference}) for both 
the pure gauge theory (open squares) and dynamical fermion simulations
(filled squares). Obviously, the fit formula 
eq.~(\ref{difference}) provides an excellent description of the data. 
We checked that also the other above-mentioned observables  can be
fitted by eq.~(\ref{difference}). We remark that for the scalar density 
the exponential growth of rounding-errors sets in for only large trajectory
lengths, $\tau\approx 8$. 
We conclude that indeed both sources of rounding-errors, the random walk
and the chaotic behaviour, seem to be present
in our simulations in practice and show up in physical
observables. 

We find the same rounding-error behaviour of eq.~(\ref{difference}) 
when we first sum up the observables
over the lattice, constructing $O_g=\sum_x O_x$, and then build the {\em global} 
difference $|O_g-\tilde{O}_g|$ at each molecular dynamics step.
The prominent exception is the global difference of the 
plaquette, $|P_g-\tilde{P}_g|$, which shows an oscillatory behaviour similar to the
ones shown in fig.~1. The reason for these oscillations is that in the HMC algorithm the
total Hamiltonian is, up to finite step size errors, conserved. 
Therefore, 
the tangent vector that corresponds to the plaquette has to compensate
the oscillations of the tangent vector that corresponds to the momenta studied in section~4.

%Let us take the 64-bit version of the program with which the ``true'' simulation
%is performed
%resulting in the true values of the observables. 
For a lattice size of $6^38$ 
we see from fig.~\ref{fig:fit} that 
the difference $\| L-\tilde{L}\|$, which indicates the deviation from the true value, 
is very small 
and hard to detect in a simulation. 
By true value we here mean --by definition-- the value of the observables from 
the 64-bit arithmetic program version. 
It is only for very long trajectories that one would risk the danger to 
produce significant differences. 

One may expect that 
the coefficients $c_i$ in eq.~(\ref{difference}) 
depend on the lattice size. 
Our original plan was therefore to 
extract the volume dependence of these coefficients and to extrapolate to a lattice
size, where rounding-error effects and in particular the consequences of irreversibility
will become dangerous for trajectories of length $1$. 
Our findings for the coefficients for $\| L-\tilde{L}\|$ 
on various lattice sizes
is given in table~\ref{table1}. To our surprise, the fit parameters show basically no lattice
size dependence at all.  A similar finding was obtained for the plaquette variable
and for plaquette correlation functions at various distances. 

\begin{table}[t]
\begin{center}
\parbox{12.5cm}{\caption{ \label{table1}
 Lattice dependence of the fit parameters in eq.~(\ref{difference})
 for the Polyakov line
 at $\beta=2.12$ for the quenched case and $\beta=2.12$, $\kappa=0.15$ for
 the unquenched simulations.
}}
\end{center}
\begin{center}
\begin{tabular}{cc*{4}{r@{.}l}}
\hline
Lattice &  & \multicolumn{2}{c}{$c_1 \cdot\, 10^6$} &
\multicolumn{2}{c}{$c_2\cdot\, 10^6$}
 & \multicolumn{2}{c}{$\nu$} & \multicolumn{2}{c}{$c_0\cdot\, 10^6$}   \\
\hline\hline
 $4^38$ & quenched  &
   0&1180(10)   &   0&108(2)    & 0&630(6)  &  $-$0&027(2)   \\
%\hline
 $6^38$ & quenched &
  0&1177(6) & 0&104(1) & 0&654(3)  &   $-$0&023(1)    \\
%\hline
 $8^4$ & quenched  &
  0&1177(6) & 0&103(1) & 0&659(3)  &   $-$0&022(1)   \\
%\hline
 $4^38$  & dynamical &
 0&1111(14) & 0&143(4) & 0&557(8)  &   $-$0&062(4)   \\
%\hline
 $6^38$  & dynamical &
 0&1066(21) & 0&143(5) &   0&608(10)   &   $-$0&060(5)   \\
%\hline
 $8^4$  & dynamical  &
 0&1115(17) & 0&134(4) & 0&623(8)  &  $-$0&052(4)   \\
\hline
\end{tabular}
\end{center}
\end{table}

%\begin{table}[t]
%\begin{center}
%
%\parbox{12.5cm}{\caption{ \label{table1}
% Lattice dependence of the fit parameters in eq.~(\ref{difference}) 
% for the Polyakov line
% at $\beta=2.12$ for the quenched case and $\beta=2.12$, $\kappa=0.15$ for 
% the unquenched simulations. 
%}}
%%
%\end{center}
%\begin{center}
%\begin{tabular}{cc*{4}{r@{.}l}}
%\hline
%Lattice &  & \multicolumn{2}{c}{$c_1 \cdot\, 10^6$} & \multicolumn{2}{c}{$c_2\cdot\,
%10^6$} 
% & \multicolumn{2}{c}{$\nu$} & \multicolumn{2}{c}{$c_0\cdot\, 10^6$}   \\
%\hline\hline
% $4^38$ & quenched  &
% \  0&1180(10) \  & \  0&108(2)  \  & 0&630(6)  & \ -0&027(2) \  \\
%%\hline
% $6^38$ & quenched &
%  0&1177(6) & 0&104(1) & 0&654(3)  &   -0&023(1)    \\
%%\hline
% $8^4$ & quenched  &
%  0&1177(6) & 0&103(1) & 0&659(3)  &   -0&022(1)   \\
%%\hline
% $4^38$  & dynamical &
% 0&1111(14) & 0&143(4) & 0&557(8)  &   -0&062(4)   \\
%%\hline 
% $6^38$  & dynamical &
% 0&1066(21) & 0&143(5) & \  0&608(10)  \ &   -0&060(5)   \\
%%\hline
% $8^4$  & dynamical  &
% 0&1115(17) & 0&134(4) & 0&623(8)  &   -0&052(4)   \\
%\hline
%\end{tabular} 
%\end{center}
%\end{table}
% 

For the HMC algorithm applied for the pure gauge theory one may give a simple reasoning:
since in the HMC algorithm the gauge links
only feel their nearest neighbours, their update is essentially local. But, for dynamical
fermions, in the equations of motion the information of the inverse fermion matrix is
used, which is {\em non-local}. 
The results of table~\ref{table1} are therefore somewhat counter-intuitive.
This indicates that the difference between the 32-bit and 64-bit versions
in the solution vectors obtained from the conjugate gradient (CG) 
method do not increase
significantly with the volume.  
We checked this explanation explicitly by studying the stability of the 
CG inversion method against rounding-errors.
In addition it might be that our studies are
performed in a situation where the propagators vanish
fast with growing distance. Therefore information would only be
transported over a few lattice spacings.

\subsection{Hamiltonian}

Would the HMC algorithm only contain the molecular dynamics
part, the results of the previous section would indicate
that rounding-error effects show up in physical observables
only for a very large number of integration steps. However,
the exactness of the HMC algorithm requires a Metropolis
reject/accept step for which the difference of the initial and final values
of the Hamiltonians are taken. An important observation is 
that for this difference 
an {\em absolute} precision 
is required. The computations of the acceptance Hamiltonians therefore play a special role and
an investigation of rounding-error effects on them is most crucial.

We therefore measured in full QCD 
the difference between the value of the Hamiltonian
$\tilde{H}$ from the 32-bit arithmetic program version and the one from the
64-bit arithmetic program version $H$. Let us define the ratio            
\begin{equation}
R_H =\frac{\langle |\delta(\Delta H)|\rangle}{\langle |\Delta H|\rangle}\; , 
\end{equation} 
where 
$\langle |\delta(\Delta H)|\rangle =\langle |\Delta\tilde{H} - \Delta H |\rangle$ and 
$\Delta H$ is the difference between the initial Hamiltonian 
at the beginning and the final
Hamiltonian evaluated at the end of a trajectory. 
One should expect a dependence on the lattice size for $R_H$.
Keeping the acceptance rate constant means that $\langle |\Delta H|\rangle$ 
should be constant. However, the values for $H$ themselves increase 
substantially with the lattice size since they are constructed
by a sum over all lattice points, see eq.~(\ref{eq:ham}). As a consequence, the rounding-errors
become more and more significant relative to $\langle |\Delta H|\rangle$.

Indeed, 
averaging over 100 trajectories, we find $R_H \approx 0.08\%, 0.18\%$ and $0.70\%$ 
for $8^4$, $12^4$ and $16^4$ lattices, respectively. 
%\begin{center}
%\begin{table}[hbt]
%\caption{We give values, for $R_H=\langle |\delta(\Delta H)|>/<|\Delta H|>$, 
%         the difference of the acceptance Hamiltonian from the 32-bit and the 64-bit arithmetic
%         program versions relative to the expectation value of the Hamiltonian obtained
%         from the 64-bit program version. }
%\vspace{2mm}
%\label{tab:table2}
%\begin{tabular}{|c|c|c|c|c|c|}
%\hline
%Lattice & $R_H$ & Lattice & $R_H$ & Lattice & $R_H$   \\
%\hline 
% $8^4$  & $0.08$ &     
% $12^4$ & $0.18$ &    
% $16^4$ & $0.70$  \\
%\hline
%\end{tabular}
%\end{table}
%\end{center} 
These numbers are obtained by performing only 10
molecular dynamics steps with step size $\delta\tau =0.02$, 
a number that is certainly too small for
simulations on a $16^4$ lattice. Still, extrapolating $R_H$ linearly in 
$N^2$ to a $32^4$ lattice
we find that $\langle |\delta(\Delta H)|\rangle$ could become about $5\%$ of 
$\langle |\Delta H|\rangle$. We think that this number is a conservative estimate
of the rounding-error effects on the acceptance Hamiltonian 
on such a large lattice. We feel that performing simulations on lattices of comparable 
sizes on 32-bit arithmetic computers 
could not be safe.
The same holds, of course, on 64-bit arithmetic machines when 32-bit arithmetics
is used for the numerical computations. 

\section{Conclusions}

In this paper, we have investigated several questions related to
the reversibility problem of the molecular dynamics kind of algorithms
as used for the simulation of lattice QCD. 
The purely classical equations of motion used in these algorithms behave like those
of a chaotic dynamical system.
We have determined the leading
Liapunov exponents $\nu$ for both pure $SU(3)$ gauge theory and full
QCD for various bare parameters. 
We have found a weak dependence of the leading Liapunov exponent on
the hopping parameter $\kappa$, whereas the $\beta$-dependence is 
more significant.
We argued that, due to
the chaotic nature of the equations of motion, the rounding-error that 
occurs in the integration of these equations is magnified 
exponentially with increasing trajectory length, when keeping the step
size constant. 

We suggested that 
the growth of the rounding-errors as a function of the trajectory
length $\tau$ should consist of two parts: the first is a random walk 
behaviour, leading to a $\sqrt{\tau}$ part. The second is the exponential
amplification, leading to an $\exp(\nu\tau )$ contribution. We verified that
this ansatz is indeed followed by observables such as the plaquette, the Polyakov
line and plaquette correlation functions. 
Considering the above observables, 
the rounding-errors, however, do not increase with the lattice
size, at least for the parameter values we considered here. 
The situation is different for the value of the difference $\Delta H$ of
the initial and final Hamiltonians, computed at the beginning and the 
end of a molecular dynamics trajectory, 
which enters the Metropolis decision in the HMC algorithm.
This quantity plays a special role in the molecular dynamics kind of algorithms,
because it is the absolute precision with which these Hamiltonians 
have to be calculated.
We find that, with growing lattice size, the rounding-errors for $\Delta H$
increase. 
Extrapolating the rounding-error effects as found on $8^4$, $12^4$ and
$16^4$ lattices to a lattice of size $32^4$, gives a rounding-error 
for $\Delta H$ that can reach several per cent. 
We regard this estimate as rather conservative and therefore 
conclude that simulations on lattices of size
$32^4$ using 32-bit arithmetics, either by hardware or by software, 
could be dangerous. 

A real test of the effects of rounding-errors on observables would be 
to run a 32-bit arithmetic program version  against a 64-bit arithmetic
version for a long time and see, whether one finds differences in some
observables. Here we could only perform approximations of this test. 
We can therefore not give a definite answer of whether rounding-errors
will lead to problems for the molecular dynamics kind of
algorithms. However, our data indicate that it can be dangerous  
to use these algorithms for simulating large lattices 
employing 32-bit arithmetics. 

\section*{Acknowledgements}

We would like to thank M.~L\"uscher for his critical
reading of the paper. We thank 
 members of the SESAM collaboration for
useful discussions. 
Critical comments on our draft by  A.~Frommer, I.~Horvath, A.~D.~Kennedy 
and R.~Sommer are
gratefully acknowledged.


\begin{thebibliography}{9}
\settowidth{\baselineskip}{\reffont QC}
%
\bibitem{tony} S. Duane, A.D. Kennedy, B.J. Pendleton and D. Roweth,
 Phys. Lett. B195 (1987) 216.
%
\bibitem{horowitz}
A.~M.~Horowitz, Phys. Lett. 156B (1985) 89; Nucl. Phys. B280 (1987) 510;
               Phys. Lett. 268B (1991) 247.
%
\bibitem{kra_us} K. Jansen and C. Liu, 
Nucl. Phys. B 453 (1995) 375 and B 459 (1996) 437.
%
\bibitem{wuppertal}
         U. Gl\"assner,
                    S. G\"usken, H. Hoeber, T. Lippert,
                    X. Luo, G. Ritzenh\"ofer, K. Schilling and G. Siegert,
                    hep-lat/9510001.
%
\bibitem{brower}
 R.~C. Brower, T. Ivanenko, A.~R. Levi and K.~N. Orginos,
Nucl. Phys. B484 (1997) 353.          
%
\bibitem{ivan}
 R.~G. Edwards, I.~Horvath, A.~D.~Kennedy,
Nucl. Phys. B484 (1997) 375.
%
\bibitem{bob} R. Edwards and A.D. Kennedy, private communications.
%
\bibitem{revus}
K.~Jansen and C.~Liu, Nucl.~Phys.~B (Proc.Suppl.) 53 (1997) 974.
%
%\bibitem{kra_boson} K.~Jansen, B.~Jegerlehner and C.~Liu, 
%Phys.Lett. {\bf B375} (1996) 255,
%
\bibitem{sugar} S. Gottlieb, W. Liu, D. Toussaint, R. L. Renken and
R. L. Sugar, Phys. Rev. D 35 (1987) 2531.
%
\bibitem{liapunov} A.~M.~Liapunov, Ann. Math. Studies 17 (1907) 1947.
%
\bibitem{oseledec} V.~I.~Oseledec, Trans. Moscow Math. Soc. 19 (1968) 197.
%
%\bibitem{pesin} Ya.~B.~Pesin, Soviet Math. Dokl. 17, 196 (1976), Russian 
%Math. Surveys 32, 55 (1977).
%
\bibitem{benettin1} G.~Benettin, L.~Galgani and J.~M.~Strelcyn, Phys. 
Rev. A14 (1976) 2338. 
%
%\bibitem{benettin2} G.~Benettin, M.~Casartelli, L.~Galgani, A.~Giorgilli
% and J.~M.~Strelcyn, Nuovo Cimento 44B, 183, Nuovo Cimento 50B 221 (1979). 
%
\bibitem{llbook} A.~J.~Lichtenberg and M.~A.~Lieberman, Regular and 
Chaotic Dynamics, 2nd ed., Springer-Verlag, 1992.
%
%\bibitem{chirikov} B.~B.~Chirikov, F.~M.~Izrailev and 
%V.~A.~Tayursky,Comput. Phys. Commun. 5, 11 (1973).

\end{thebibliography}
\end{document}